\pdfoutput=1
\documentclass[cernpreprint,atlasdraft=true,texlive=2016,UKenglish,texmf]{atlasdoc}
 
\usepackage{atlaspackage}
\usepackage{atlasbiblatex}
 
\usepackage{atlasphysics}
 
\graphicspath{{logos/}{figures/}}
 
\usepackage{ANA-HIGG-2018-27-PAPER-defs}

\usepackage[flushleft]{threeparttable}
\usepackage{comment}
 
\AtlasTitle{Search for resonances decaying into photon pairs in 139~fb$^{-1}$ of $pp$ collisions at $\sqrt{s} = 13$~TeV with the ATLAS detector}
 
\AtlasAbstract{
Searches for new resonances in the diphoton final state, with spin 0 as predicted by theories with an extended Higgs sector and with spin 2 using a warped extra-dimension benchmark model, are presented using 139~fb$^{-1}$ of $\sqrt{s} = 13$~TeV $pp$ collision data collected by the ATLAS experiment at the LHC. No significant deviation from the Standard Model is observed and upper limits are placed on the production cross-section times branching ratio to two photons as a function of the resonance mass.
}
 
\author{The ATLAS Collaboration}
 
\AtlasRefCode{HIGG-2018-27}
\PreprintIdNumber{CERN-EP-2020-248}

\AtlasJournalRef{Phys. Lett. B 822 (2021) 136651}
\AtlasDOI{10.1016/j.physletb.2021.136651}
 
\hypersetup{pdftitle={ATLAS document},pdfauthor={The ATLAS Collaboration}}
 
\begin{document}

\newcounter{num}
\setcounter{num}{1} 
 
\maketitle

\ifnum\value{num}>0 {\section{Introduction}
} \else {}\fi
 
Many theories of physics beyond the Standard Model (SM) predict the presence of new high-mass states which can currently only be produced in the high-energy collisions at the Large Hadron Collider (LHC).
One powerful experimental signature of these new states is a resonance in the diphoton invariant mass spectrum, as this final state provides excellent invariant mass resolution which can be used to separate the signal from the SM background processes.
 
This Letter presents a search for new high-mass resonances decaying into two photons using 139~\ifb\ of $\sqrt{s}=13$~\TeV\ proton--proton ($pp$) collision data recorded by the ATLAS detector from 2015 to 2018 at the LHC.
This analysis searches for a generic resonance using two benchmark signal models: a spin-0 resonant state ($X$), which is predicted by many models that include extensions to the Higgs sector~\cite{Lee:1973iz,Hill:1987ea,Veltman:1989vw,Binoth:1996au,Schabinger:2005ei,Patt:2006fw,Pruna:2013bma}; and a spin-2 graviton ($G^{*}$), taken here to be the lightest Kaluza--Klein (KK)~\cite{Appelquist:1987nr} excitation in a Randall--Sundrum model~\cite{Randall:1999ee,Randall:1999vf} with one warped extra dimension (RS1).
Previous searches for high-mass diphoton resonances using the 2015--2016 $pp$ collision data have been reported by the ATLAS and CMS collaborations~\cite{HIGG-2016-17,CMS-EXO-17-017}.
Modifications with respect to the analysis in Ref.~\cite{HIGG-2016-17} include a common event selection for the spin-0 and spin-2 resonance searches, the use of the functional decomposition method~\cite{Edgar:2018irz} to assess the spurious-signal uncertainty, and updates to the photon reconstruction, identification, isolation and energy calibration.
 
This search uses functional forms to describe the signal and background components in a fit of the diphoton invariant mass spectrum, $m_{\gamma\gamma}$, to determine the signal yield.
In the absence of a significant signal excess, limits are placed on the production cross-section times branching ratio, $\sigma \times B(X\rightarrow \gamma\gamma)$. For the spin-0 case, the limits are computed in terms of the fiducial cross-section, defined as the product of the cross-section times the branching ratio to two photons within a fiducial acceptance which closely follows the selection criteria applied to the reconstructed data. In the spin-2 analysis, limits are instead placed on the total cross-section times branching ratio to the two-photon final state.
These limits are provided as a function of resonance mass and width in the spin-0 search and as a function of graviton mass and coupling \kMpl\ for the spin-2 resonance search.
As the graviton natural width is related to the coupling via $\Gamma_{G^{*}}=1.44(\kMpl)^2\mG$, the graviton is expected to be a narrow resonance~\cite{Davoudiasl:1999jd} for the couplings considered in this search, i.e.\ \kMpl\ < 0.1.


\ifnum\value{num}>0 {
\section{ATLAS detector}
} \else {}\fi

\newcommand{\AtlasCoordFootnote}{
ATLAS uses a right-handed coordinate system with its origin at the nominal interaction point (IP)
in the centre of the detector and the \(z\)-axis along the beam pipe.
The \(x\)-axis points from the IP to the centre of the LHC ring,
and the \(y\)-axis points upwards.
Cylindrical coordinates \((r,\phi)\) are used in the transverse plane,
\(\phi\) being the azimuthal angle around the \(z\)-axis.
The pseudorapidity is defined in terms of the polar angle \(\theta\) as \(\eta = -\ln \tan(\theta/2)\).
Angular distance is measured in units of \(\Delta R \equiv \sqrt{(\Delta\eta)^{2} + (\Delta\phi)^{2}}\).}
 
The ATLAS experiment~\cite{PERF-2007-01, ATLAS-TDR-2010-19, PIX-2018-001} at the LHC is a multipurpose particle detector
with a forward--backward symmetric cylindrical geometry and a near \(4\pi\) coverage in
solid angle.\footnote{\AtlasCoordFootnote}
It consists of an inner tracking detector surrounded by a thin superconducting solenoid
providing a \SI{2}{\tesla} axial magnetic field, electromagnetic and hadronic calorimeters, and a muon spectrometer.
The inner tracking detector covers the pseudorapidity range \(|\eta| < 2.5\).
It consists of silicon pixel, silicon microstrip, and transition radiation tracking detectors.
Lead/liquid-argon (LAr) sampling calorimeters provide electromagnetic (EM) energy measurements
with high granularity.
A hadron sampling calorimeter made of steel and scintillator tiles covers the central pseudorapidity range (\(|\eta| < 1.7\)).
The endcap and forward regions are instrumented with LAr calorimeters
for EM and hadronic energy measurements up to \(|\eta| = 4.9\).
The muon spectrometer surrounds the calorimeters and is based on
three large air-core toroidal superconducting magnets with eight coils each.
The field integral of the toroids ranges between \num{2.0} and \SI{6.0}{\tesla\metre}
across most of the detector.
A two-level trigger system is used to select events.
The first-level trigger is implemented in hardware and uses a subset of the detector information
to keep the accepted rate below \SI{100}{\kHz}.
A software-based trigger reduces the accepted event rate
to an average of \SI{1}{\kHz} depending on the data-taking conditions.
 

\ifnum\value{num}>0 {\section{Data and simulated event samples}
} \else {}\fi

The search is carried out using the $\sqrt{s} = 13$~\TeV\ $pp$ collision dataset with a bunch spacing of \SI{25}{\ns} collected from 2015 to 2018, with stable beam conditions and all ATLAS subsystems operational, which corresponds to an integrated luminosity of $139.0\pm2.4~\ifb$~\cite{ATLAS-CONF-2019-021}.
The data were recorded using a diphoton trigger that required two electromagnetic clusters with transverse energies \ET\ above 35 and 25~\GeV, respectively, both fulfilling photon identification criteria based on shower shapes in the electromagnetic calorimeter.
The efficiency of the diphoton trigger with respect to the full event selection is above 99\% for 2015--2016 and above 98\% for 2017--2018~\cite{TRIG-2018-05}.
 
Simulated Monte Carlo (MC) events are used to optimize the analysis selections and characterize the signal and background. 
Interference effects between the resonant signal and all background processes are expected to be small for a narrow-width signal and are neglected in this analysis to keep it as model-independent as possible.
 
To model the spin-0 resonance signals, MC samples were generated for a hypothetical resonance mass \mX\ in the range 200--3000~\GeV\ assuming a decay width \GX\ of 4~\MeV\ to describe a hypothetical resonance in the narrow-width approximation (NWA). Four production modes were generated to assess the model dependence of the fiducial selection.
The gluon--gluon fusion (ggF) samples were generated at next-to-leading order (NLO) in pQCD using \MGNLO~\cite{Alwall:2014hca}, with a set-up matching that described in Ref.~\cite{HIGG-2016-17}.
Vector-boson fusion (VBF) samples were generated at NLO in pQCD with \POWHEGBOX v2~\cite{Nason:2004rx,Frixione:2007vw,Alioli:2010xd} interfaced with \PYTHIA[8]~\cite{Sjostrand:2014zea} using the CT10 parton distribution function (PDF) set~\cite{Lai:2010vv} and the AZNLO tune~\cite{STDM-2012-23} of the parameter values.
Simulated samples for production of a scalar resonance in association with a vector boson or $t\bar{t}$ pair were generated at leading order (LO) in pQCD with \PYTHIA[8] using the NNPDF23LO PDF set~\cite{Ball:2012cx} and the A14 tune~\cite{ATL-PHYS-PUB-2014-021}.
Additional gluon--gluon fusion samples were produced for various fixed decay widths, ranging from 4~\MeV\ up to 10\% of \mX.
 
To model the spin-2 resonance signals, MC samples incorporating the RS1 model were generated at LO in pQCD using \PYTHIA[8] with the NNPDF23LO PDF set and the A14 tune, with masses \mG\ in the range 500--5000~\GeV\ and fixed coupling values \kMpl\ ranging from 0.01 to 0.1.

The irreducible background in this search, coming from events with two prompt photons, was simulated using the \SHERPA~\cite{Gleisberg:2008ta,Bothmann:2019yzt} event generator, version 2.2.4.
Matrix elements were calculated with up to one additional parton at NLO and with two or three partons at LO in pQCD and merged with the \SHERPA parton-shower simulation using the ME+PS@NLO prescription~\cite{Hoeche:2011fd,Hoeche:2012yf,Catani:2001cc,Hoeche:2009rj}.
The NNPDF3.0 NNLO PDF set~\cite{Ball:2014uwa} was used in conjunction with a dedicated parton-shower tune in the \SHERPA generator.
 
The effects of multiple $pp$ interactions in the same bunch crossing as the hard scatter and in neighbouring ones are included using simulated events generated with \PYTHIA[8]. Simulated events were weighted to reproduce the distribution of the average number of interactions per bunch crossing observed in data.
 
All simulated signal events were processed using a full simulation of the ATLAS detector~\cite{SOFT-2010-01} based on $\GEANT$~\cite{Agostinelli:2002hh}.
The background $\gamma\gamma$ events were processed using a fast simulation of the ATLAS detector~\cite{ATL-PHYS-PUB-2010-013}, where the full simulation of the calorimeter is replaced with a parameterization of the calorimeter response.
All simulated events were reconstructed with the same reconstruction algorithms as those used for data.

\ifnum\value{num}>0 {\section{Object and event selection}
} \else {}\fi

Photon candidates are reconstructed from topological clusters of energy deposits in the EM calorimeter and calibrated as described in Ref.~\cite{EGAM-2018-01}.
The event selection requires at least two photon candidates with $\ET> 25$~\GeV\ and $|\eta|< 2.37$, excluding the barrel-to-endcap transition regions of the calorimeter, $1.37 <|\eta|< 1.52$. The two highest-\ET\ photons and additional information from the tracking systems are used to identify the diphoton production vertex~\cite{HIGG-2013-08}.
 
To reduce the background from jets, photon candidates are required to fulfil \emph{tight} identification criteria based on shower shapes in the EM calorimeter and energy leakage into the hadronic calorimeter~\cite{EGAM-2018-01}. A looser identification is considered for background estimations that benefit from a larger sample size. The \emph{tight} identification is optimized in sub-ranges of photon \ET\ and $|\eta|$ to have an identification efficiency greater than 90\% for the \ET\ range considered in this analysis.
The calorimeter isolation transverse energy \ETiso\ is required to be smaller than $0.022\ET + 2.45$~\gev, where \ETiso\ is defined as the sum of transverse energies of the positive-energy topological clusters~\cite{PERF-2014-07} within a cone of size $\Delta R = 0.4$ around the photon candidate. The core of the photon shower is excluded, and \ETiso\ is corrected for the leakage of the photon shower into the isolation cone. The contributions from the underlying event and pile-up are subtracted using the techniques described in Refs.~\cite{Cacciari:area,Cacciari:UE}. The track isolation is defined as the scalar sum of the transverse momenta of tracks with $\pT > 1$~\GeV\ which satisfy some loose track-quality criteria, are not associated with a photon conversion and originate from the diphoton production vertex in a $\Delta R=0.2$ cone around the photon candidate. It is required to be less than $0.05\ET$.
 
The diphoton invariant mass \mgg\ is evaluated using the energies of the leading- and subleading-\ET\ photons, their separation $\Delta\phi$ in azimuthal angle and $\Delta\eta$ in pseudorapidity determined from their positions in the calorimeter, and the position of the diphoton production vertex.
Additional kinematic requirements are placed on the photon \ET\ relative to the diphoton invariant mass: the leading photon must have $\ET/\mgg>0.3$ and the subleading photon must satisfy $\ET/\mgg>0.25$.
These requirements were optimized to retain improved significance with respect to Ref.~\cite{HIGG-2016-17} while allowing the analysis to have a unified selection for the spin-0 and spin-2 models.
For the spin-0 model, the new selections result in a slightly worse expected limit for the NWA scalar models below 700~\GeV\ but up to 30\% better expected limits above.
For the spin-2 model, these updated selections lead to an improvement in the expected limit by 15\% to 50\% depending on \mG\ for masses up to 2 TeV.
 
In total, 433\,919 data events with $\mgg > 150$~\GeV\ are selected.
 
The fiducial volume for the spin-0 interpretation is defined by requiring two photons at generator level with $|\eta|<2.37$, and $\ET/\mgg>0.3$ and $\ET/\mgg>0.25$ for the leading and subleading photons, respectively. The particle isolation, defined as the scalar sum of \pT\ of all the stable particles (except neutrinos) found within a $\Delta R=0.4$ cone around the photon direction, is required to be less than $0.05\ET + 6$~\GeV.


\ifnum\value{num}>0 {\section{Signal modelling}
}
\else {}\fi
 
Parametric models of the diphoton invariant mass distributions of the signal are used to test for different resonance masses and \GX\ or \kMpl. The detector resolution is modelled by a double-sided Crystal Ball (DSCB) function~\cite{HIGG-2014-04, Oreglia:1980cs}, with parameters expressed as a function of \mX. 
The DSCB function is convolved with the true lineshape, which is parameterized as the product of a relativistic Breit--Wigner (BW) function and mass-dependent factors accounting for the parton luminosity and the matrix elements of the processes. This convolution is performed independently for the spin-0 and spin-2 models.
The lineshapes for the spin-0 model are derived from the functional forms defined in the \MGNLO\ MC simulation of the ggF process, where a BW function of width \GX\ is combined with a mass-dependent gluon--gluon luminosity functional form.
The lineshapes for the spin-2 model are derived from the functional forms defined in the \PYTHIA[8] MC simulation of the RS1 model~\cite{Bijnens:2001gh}, where a BW function of width $1.44(\kMpl)^{2}\mG$ is combined with mass-dependent gluon--gluon and quark--antiquark luminosity functional forms.


\ifnum\value{num}>0 {\section{Background estimates}
} \else {}\fi

The unified spin-0 and spin-2 search selections in this analysis allow the use of a common background modelling procedure.
The largest background component comes from the non-resonant production of photon pairs ($\gamma \gamma$ events); smaller backgrounds come from events containing a photon and a jet ($\gamma j$ events) and events with two jets ($jj$ events), where the jets are misidentified as photons.
The relative contribution of these processes is determined using the two-dimensional sideband method described in Ref.~\cite{STDM-2012-05} and shown in Figure~\ref{fig:decomp_result}.
The overall $\gamma \gamma$ purity of the selected events increases with \mgg\ from around 89\% at 150--200~\GeV\ to around 97\% above 400~\GeV\ with an uncertainty of 0.5--3\%.
No significant difference in purity is seen between the LHC data-taking periods, and the remaining background is dominated by $\gamma j$ events.
 
\begin{figure}[hbpt]
\centering
\includegraphics[width=0.8\columnwidth]{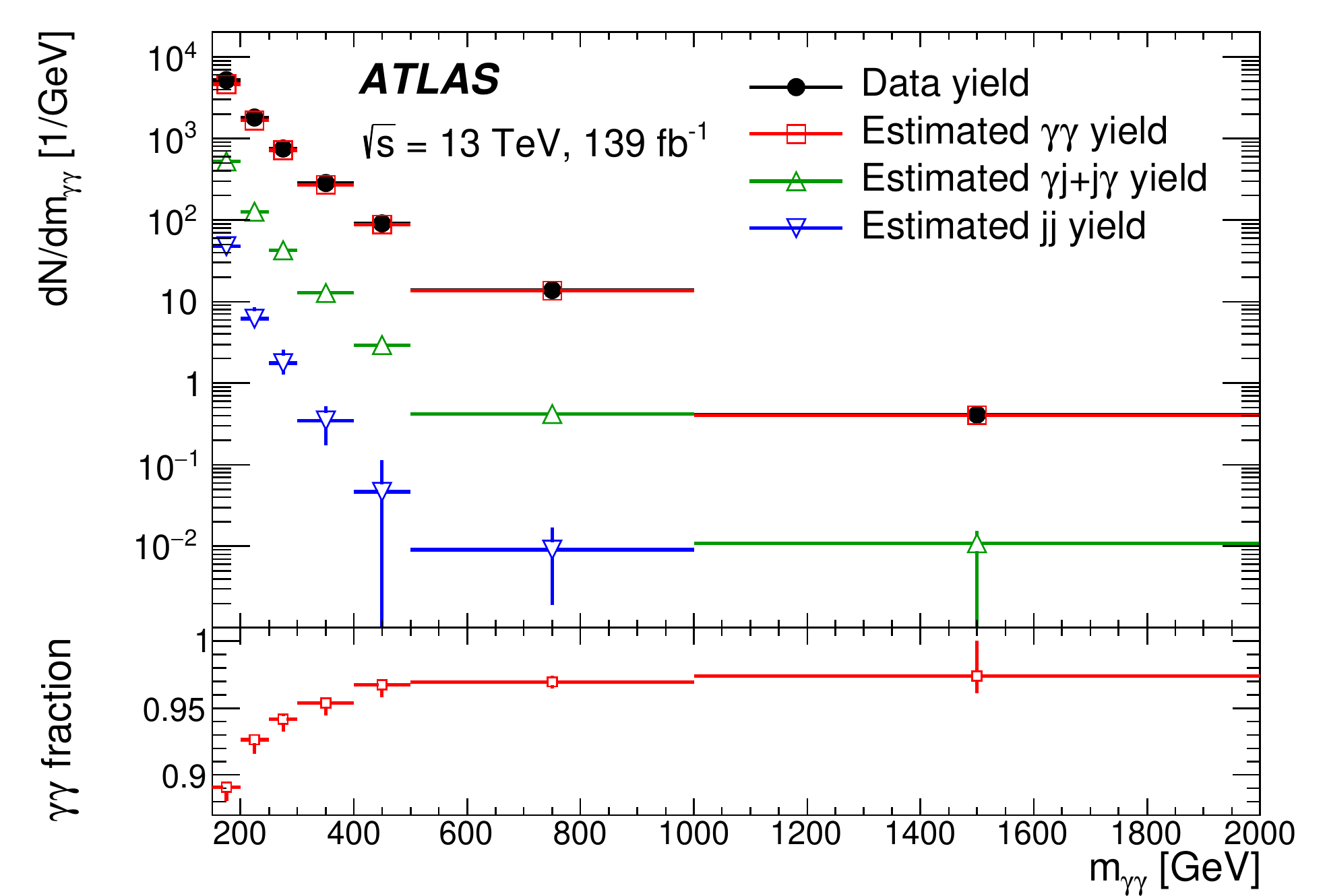}
\caption{Diphoton invariant mass distributions of the data after event selection and their decomposition into contributions from genuine diphoton ($\gamma\gamma$), photon+jet ($\gamma j$ and $j\gamma$) and dijet ($jj$) events as determined using the two-dimensional sideband method. The bottom panel shows the sample's purity in diphoton events. Each point in the distributions is plotted in the centre of the corresponding bin. The total uncertainties, including statistical and systematic components added in quadrature, are shown as error bars.}
\label{fig:decomp_result}
\end{figure}

The analytic form of the continuum background is determined by fitting a template built from reconstructed \SHERPA $\gamma\gamma$ MC events and $\gamma j$ events derived from a dedicated data control region where the photon identification requirements are inverted.
The diphoton invariant mass distribution is fitted in the range above 150~\GeV. The lower value of the range is chosen to allow for enough events in the \mgg\ side-bands to ensure a good description by the analytic form, while the upper value of the range is chosen such as the procedure remains stable in injection tests .
The search region for a resonant signal covers the region 160--3000~\GeV\ for a NWA spin-0 resonance, 400--2800~\GeV\ for a wide spin-0 resonance, and 500--2800~\GeV\ for a graviton resonance.
The procedure described in Ref.~\cite{HIGG-2013-08} is used to ensure that the chosen analytic function is flexible enough to model any potential variations in the background template. The analytic form is chosen from the family of functions previously used to describe the diphoton invariant mass spectrum~\cite{HIGG-2016-08}. Additional background templates are constructed from variations due to the uncertainties in the measured background composition, shape of the $\gamma j$ component and theoretical uncertainties such as the choice of PDF and the variation of renormalization and factorization scales.
The function with the fewest degrees of freedom that maintains enough flexibility to model all variations of the template shape is chosen, and it takes the form
\begin{eqnarray}
f(x;b,a_0,a_1) = N (1-x^{1/3})^{b} x^{a_0 + a_1 \textrm{log}(x)}
\label{eqn:powlog1}
\end{eqnarray}
where $N$, $b$, $a_0$, $a_1$ are free parameters and $x = \mgg/\sqrt{s}$.
 
Fits to these background templates are used to estimate the bias due to this choice of analytic function; any fitted signal yield is considered as a `spurious-signal' systematic uncertainty.
Up to masses of 2~\TeV\ this uncertainty is dominated by the statistical fluctuations of the background template due to the limited number of generated MC events, leading to fluctuations that are significant compared to the expected statistical uncertainty on the data.
To suppress the impact of these fluctuations, each background template is smoothed using the functional decomposition (FD) method~\cite{Edgar:2018irz}. This method uses a linear combination of orthonormal exponential functions to fit a background template and the fit result is binned and used as a smoothed template for the spurious-signal determination only.
 
To illustrate the FD process, an example fit to a pseudo-experiment and the resulting spurious-signal uncertainty is shown in Figure~\ref{fig:fd_pseudoexperiment}.
This pseudo-experiment dataset is generated from the functional form given in Eq.~(\ref{eqn:powlog1}), where the parameters are determined by fitting simulated $\gamma\gamma$ events.
First, the FD method is used to fit the pseudo-experiment, resulting in a smoothed FD template; this template and the pseudo-experiment are shown in Fig.~\ref{fig:fd_pexp}.
The spurious-signal estimated by fitting the smoothed FD template with the signal-plus-background model has fewer fluctuations and a smaller amplitude than the one estimated by fitting the unsmoothed template as shown in Fig.~\ref{fig:fd_comp}: it can be larger than 50\% of the statistical uncertainty with the unsmoothed template while it decreases from 40\% at 160~\GeV\ to 20\% at 500~\GeV\ and less than 10\% above 1~\TeV\ with the smoothed template.
To verify that the smoothing maintains the underlying shape of the $m_{\gamma\gamma}$ distribution, this process is repeated for an ensemble of pseudo-experiments and the mean number of extracted signal events is compared between the smoothed and unsmoothed templates. The resulting bias in the determination of the spurious-signal uncertainty, shown in Fig.~\ref{fig:fd_ensemble}, agrees between the unsmoothed and smoothed template ensembles and is found to be much smaller than the uncertainty from a single pseudo-experiment. It should be noted that the mean spurious-signal is not exactly equal to 0 because of the limited number of pseudo-experiments.
Finally, by using this smoothed template approach, an improved description of the spurious-signal uncertainty is obtained, leading to an improvement of 2--28\% in the expected sensitivity of this search as shown in Fig.~\ref{fig:fd_fidlim}.
 
\begin{figure}[hbpt]
\centering
\subfloat[]{\includegraphics[width=0.48\columnwidth]{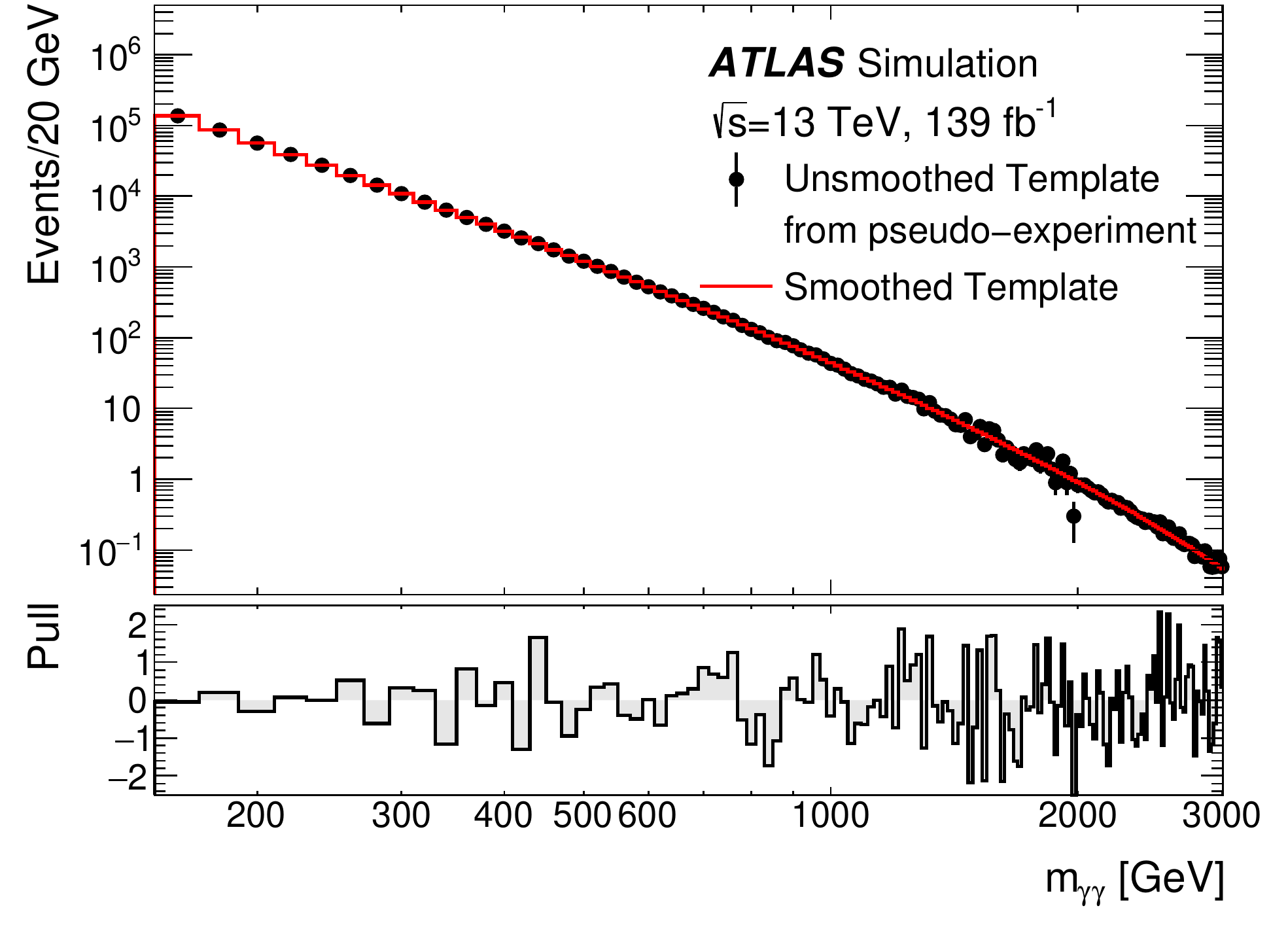}\label{fig:fd_pexp}}
\subfloat[]{\includegraphics[width=0.48\columnwidth]{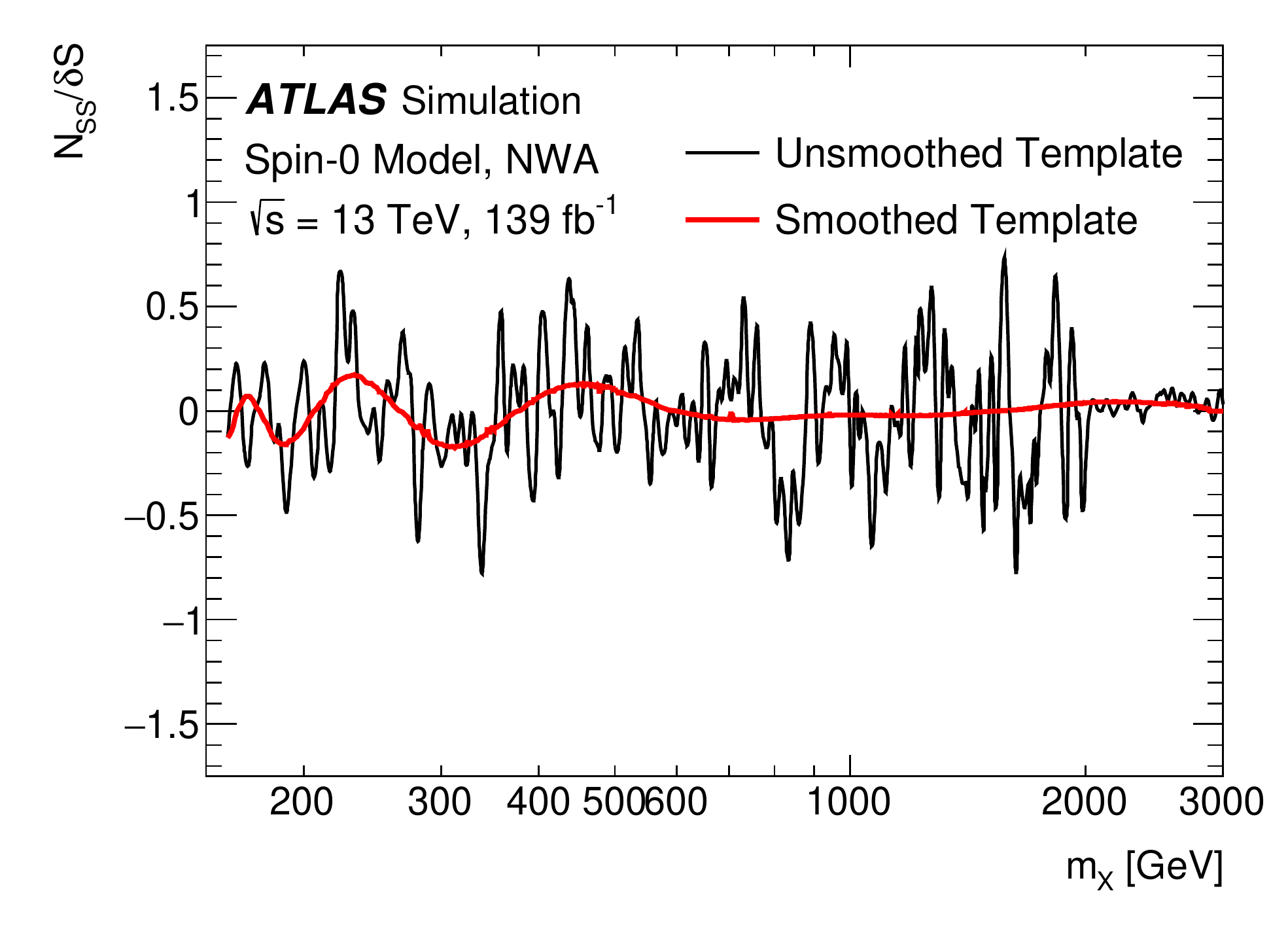}\label{fig:fd_comp}} \\
\subfloat[]{\includegraphics[width=0.48\columnwidth]{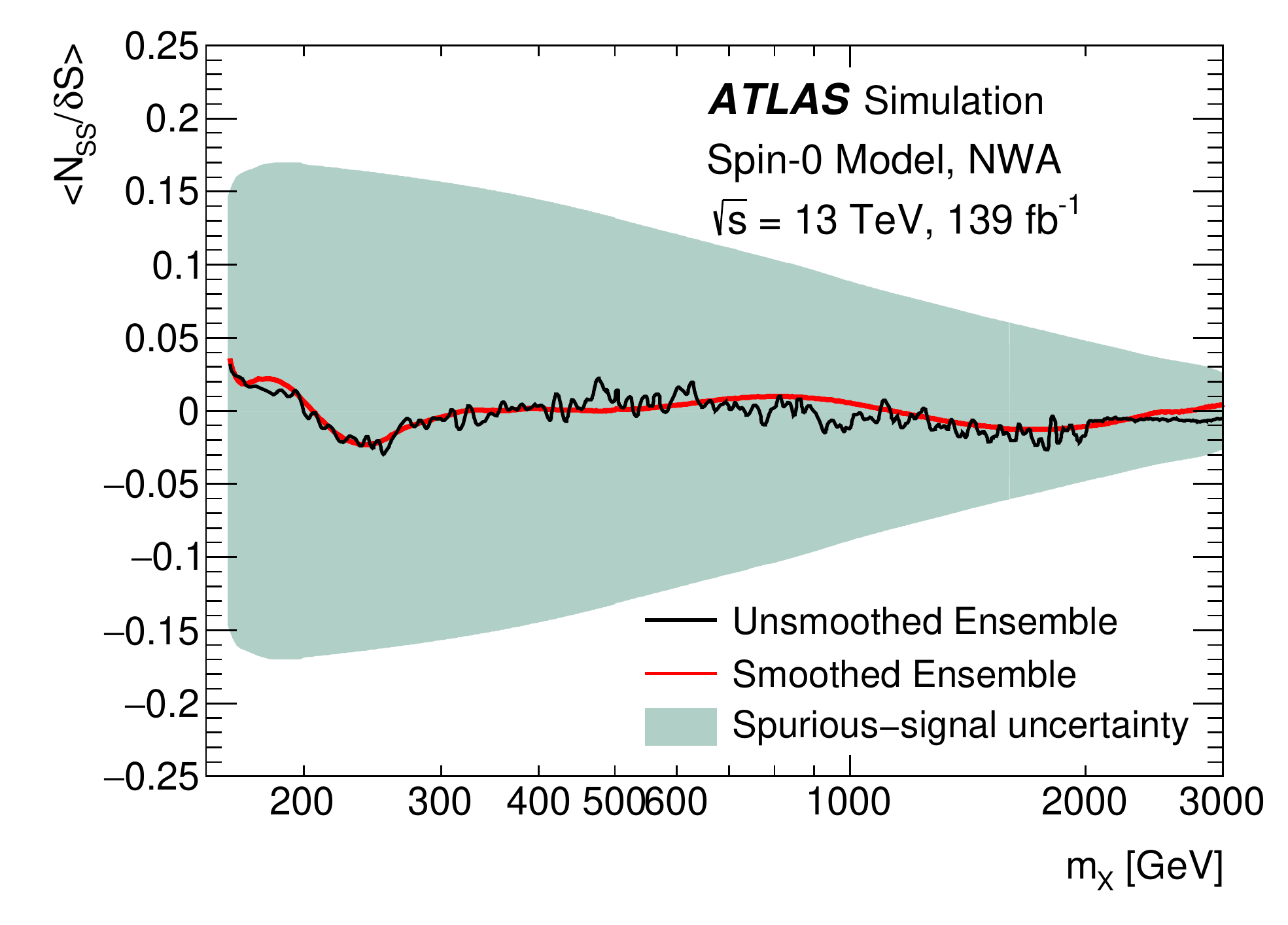}\label{fig:fd_ensemble}}
\subfloat[]{\includegraphics[width=0.48\columnwidth]{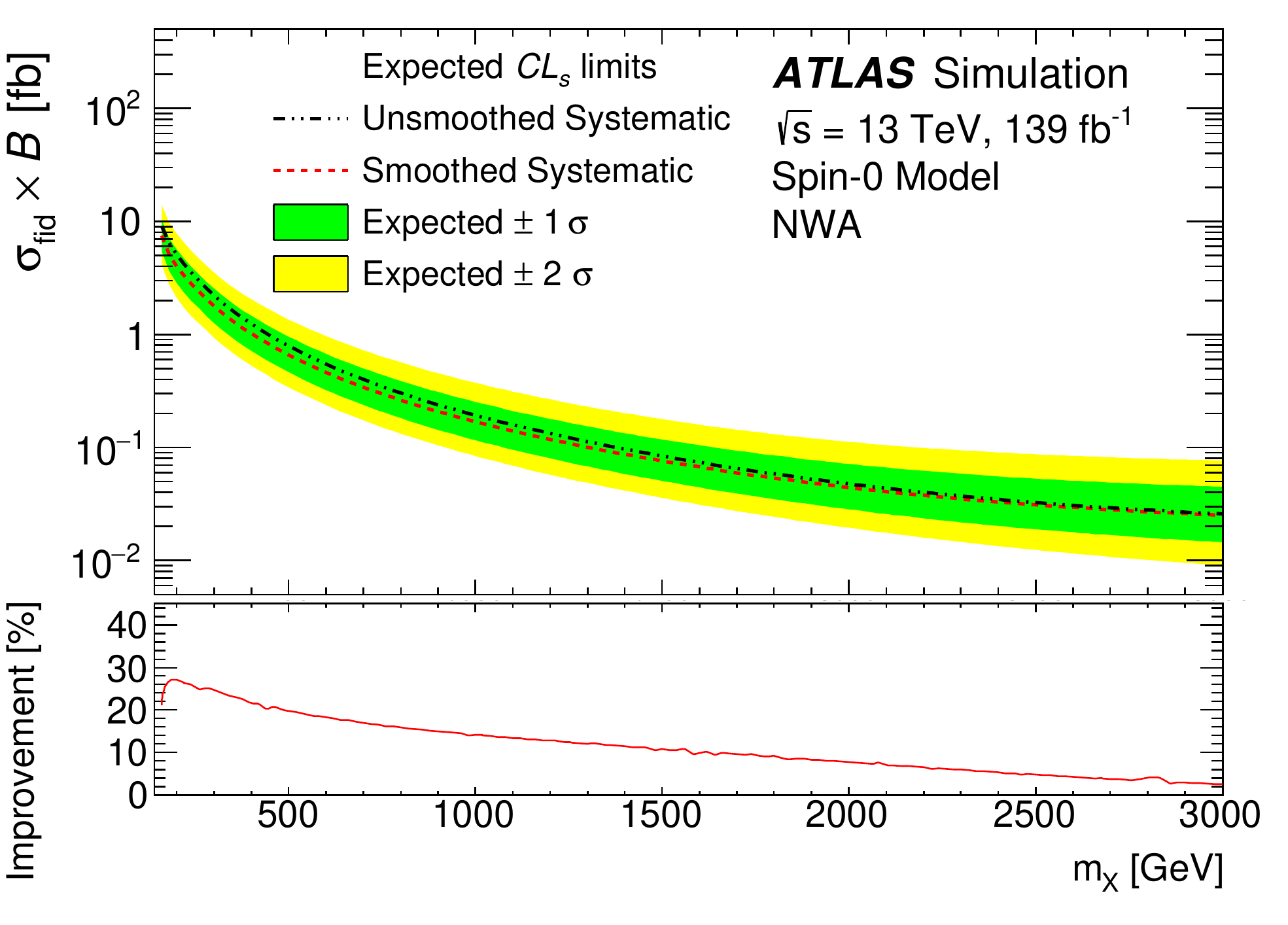}\label{fig:fd_fidlim}}
\caption{An illustration of the smoothing method used to determine the spurious-signal systematic uncertainty.
(a) An example FD fit to the $\mgg$ spectrum of a pseudo-experiment dataset generated from the functional form fit to simulated $\gamma\gamma$ events.
(b) Number of spurious-signal events relative to the statistical uncertainty of the input dataset calculated using signal-plus-background fits to the input dataset (black) and the smoothed FD template (red).
The background function used in these fits is the same function used to generate the pseudo-experiment dataset, and the variance of the unsmoothed template highlights the role of the number of events in the MC sample in the systematic uncertainty determination.
(c) Mean of the relative spurious-signal calculated by repeating the signal-plus-background fits on a large ensemble of pseudo-experiments and the FD smoothed templates. The shaded area indicates the size of the spurious-signal uncertainty from the single smoothed template pseudo-experiment of Fig.~\ref{fig:fd_comp}.
(d) The effect of the reduced systematic uncertainty on the expected fiducial cross-section limits for the NWA spin-0 signal hypothesis. The use of smoothed templates leads to a limit improvement of 2--28\%.}
\label{fig:fd_pseudoexperiment}
\end{figure}
 
The spurious-signal uncertainty for this background function choice is 40--10\% of the statistical uncertainty of the fitted signal yield for NWA signals with masses ranging from 160~\GeV\ to 3000~\GeV.
This uncertainty increases to 70--20\% for wider signals with masses ranging from 400~\GeV\ to 2800~\GeV.


\ifnum\value{num}>0 {\section{Statistical procedure}
} \else {}\fi

The statistical analysis of the data uses binned maximum-likelihood fits of the \mgg\ distribution. The systematic uncertainties listed in Table~\ref{tab:systs} are accounted for in the fits by using nuisance parameters constrained by Gaussian penalty terms in the likelihood function.
They include the experimental uncertainties in the luminosity determination, pile-up profile, trigger efficiency, and photon identification and reconstruction.
Additional systematic uncertainties in the signal shape come from the uncertainties in the photon energy resolution and scale.
For the spin-0 model, the correction factors inside the fiducial volume, used to compute the limit on the fiducial cross-section, are determined using gluon–gluon fusion MC events. They are compared with the ones computed for the other production modes, leading to a small mass-dependent systematic uncertainty in the signal yield.
 
A local $p$-value ($p_{0}$) for the background-only hypothesis is calculated using the asymptotic approximation~\cite{Cowan:2010js}. Global significance values are computed using background-only pseudo-experiments to account for the trial factors because the searches scan both the signal mass and a range of width (coupling) hypotheses. For these calculations, a maximum-likelihood fit is performed on each pseudo-experiment with the signal mass, width, and signal rate as free parameters within a specific search range. For the spin-0 case, the search ranges are defined to be $\mX = [400,\,2800]$~\GeV\ and $\Gamma_{X}/m_{X} = [0,\,10] \%$; the spin-2 case is restricted to the ranges $\mG = [500,\,2800]$~\GeV\ and $\kMpl = [0.01,\,0.1]$.
 
In the absence of a signal, the expected and observed 95\% confidence level (CL) exclusion limits on the cross-section times branching ratio to two photons are computed using a modified frequentist approach CL$_{\textrm{s}}$~\cite{Junk:1999kv,Read:2002hq}, assuming the asymptotic approximation for the test-statistic distribution. In regions where very few events are observed and the expected signal yield is also small, the asymptotic approximation is found to deviate from the limits based on pseudo-experiments. 
To account for this effect, for masses greater than 1000~\GeV, pseudo-experiments are used to verify the expected and observed limits, and used in place of the asymptotic limit where differences are observed.

For the spin-2 model, the limits on the total production cross-section can be translated into a lower limit on the graviton mass. For that purpose the spin-2 exclusion limits are extended up to $\mG = 5000$~\GeV\ for all coupling hypotheses.

\begin{table}
\centering
\caption{Summary of the systematic uncertainties. The mass-dependent values are highlighted with a star $\ast$ and are given for signal masses between 160 and 2800~\GeV\ unless stated otherwise.}
\label{tab:systs}
\vspace{0.2cm}
\begin{threeparttable}
\def\arraystretch{1.2}
\begin{tabular}{lclc}
\hline\hline
\multicolumn{4}{c}{\em Signal yield}  \\
\hline
Luminosity               & $\pm 1.7 \%$~~~ & Trigger & $\pm 0.5  \%$~~~ \\
Photon identification    & $\pm 0.5 \%$~~~ & Photon isolation & $\pm 1.5 \%$~~~ \\
Photon energy scale/resolution      & negligible & Pile-up reweighting\tnote{$\ast$} & ~~~$\pm$(2--0.2)\%  \\
Spin-0 production process\tnote{$\ast$}  & $\pm$(7--3)\%  & & \\
\hline
\multicolumn{4}{c}{\em Signal modelling\tnote{$\ast$}}                                                                                \\
\hline
Photon energy resolution & $^{+14\%}_{-9.3\%}$ -- $^{+51\%}_{-29\%}$ & &  \\
Photon energy scale & $\pm$(0.5--0.6)\% & & \\
Pile-up reweighting  & negligible & &  \\
\hline
\multicolumn{4}{c}{\em Spurious-signal, Spin-0$^{\ast}$}                                                                                      \\
\hline
NWA                  & \multicolumn{3}{l}{114--0.04 events ($\mX=160$--2800~\gev)} \\
$\Gamma_X/\mX = 2\%$  & \multicolumn{3}{l}{107--0.14 events ($\mX=400$--2800~\gev)} \\
$\Gamma_X/\mX = 6\%$  & \multicolumn{3}{l}{223--0.38 events ($\mX=400$--2800~\gev)} \\
$\Gamma_X/\mX = 10\%$ & \multicolumn{3}{l}{437--0.50 events ($\mX=400$--2800~\gev)} \\
\hline
\multicolumn{4}{c}{\em Spurious-signal, Spin-2$^{\ast}$}                                                                                      \\
\hline
$\kMpl = 0.01$       & \multicolumn{3}{l}{4.71--0.04 events ($\mG=500$--2800~\gev)}  \\
$\kMpl = 0.05$       & \multicolumn{3}{l}{19.0--0.09 events ($\mG=500$--2800~\gev)}  \\
$\kMpl = 0.1$        & \multicolumn{3}{l}{31.2--0.20 events ($\mG=500$--2800~\gev)}  \\
\hline \hline
\end{tabular}
\begin{tablenotes}
\item[$\ast$] mass-dependent
\end{tablenotes}
\end{threeparttable}
\end{table}


\ifnum\value{num}>0 {\section{Results}
} \else {}\fi

The diphoton invariant mass distribution of the events passing the selection is shown in Figure~\ref{fig:BkgOnlyFitData}, along with the background-only fit. The highest-mass diphoton candidate selected in data has a mass of 2.36~\TeV.
 
\begin{figure}[h]
\centering
\includegraphics[width=.60\textwidth]{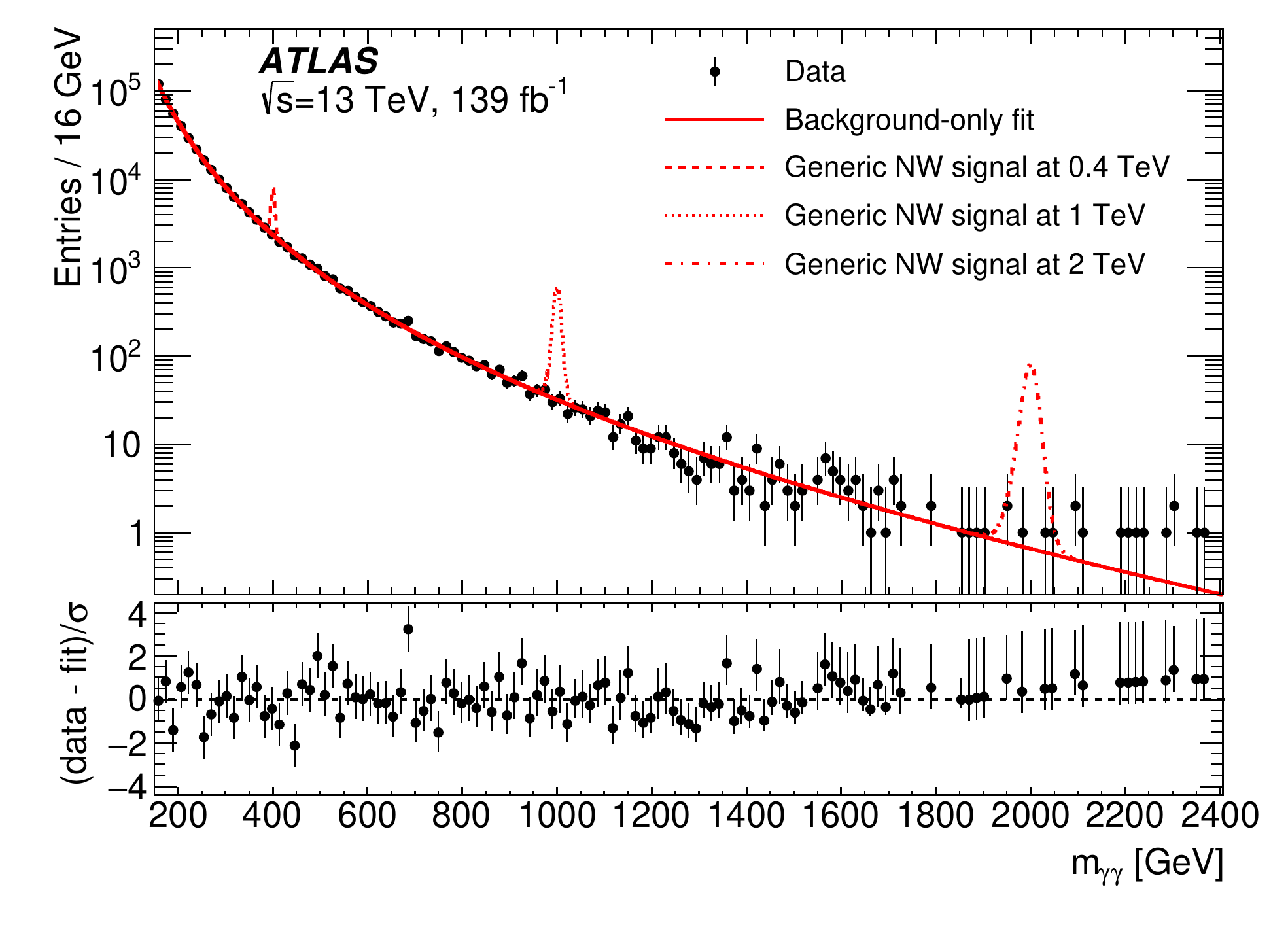}
\caption{Background-only fit to the data (black dots) as a function of the diphoton invariant mass \mgg with several generic narrow-width signal shapes overlaid (dotted lines), scaled to 10 times the value of the corresponding expected upper limit at 95\% CL on the fiducial cross-section times branching ratio, with pole masses of $\mX = 0.4$, 1 and 2~\TeV. The normalized residuals between the data and the fit are shown in the bottom panel with $\sigma$ denoting only the statistical error of the data. There is no data event with $\mgg > 2.36$~\TeV. }
\label{fig:BkgOnlyFitData}
\end{figure}
 
The probability of compatibility with the background-only hypothesis, quantified with the local $p$-value expressed in standard deviations, is shown in Figure~\ref{fig:2Dp0} as a function of the hypothesized resonance mass and width.
The most significant excess is observed for a mass of $\simeq 684~\GeV$ for the spin-0 NWA model and for the spin-2 $\kMpl=0.01$ model, corresponding to a $3.29\,\sigma$ local significance. The corresponding global significance is $1.30\,\sigma$ and $1.36\,\sigma$ for the two models, respectively. 
 
The observed and expected limits, shown in Figure~\ref{fig:1Dlimits} for two values of the signal width, are in good agreement, consistent with the absence of a signal.
 
The observed limits for different masses and different values of $\Gamma_X/\mX$ or \kMpl\ are summarized in Table~\ref{tab:results}.
For all signals considered in this analysis, the largest single improvement in the limits with respect to Ref.~\cite{HIGG-2016-17} comes from the increased integrated luminosity.
The RS1 model is excluded for \mG\ below 2.2, 3.9 and 4.5~\TeV\ for \kMpl\ values of 0.01, 0.05 and 0.1, respectively.

\begin{table}
\centering
\caption{Summary of the observed upper limits on the fiducial and the total production cross-section times branching ratio to two photons for the spin-0 and spin-2 models, respectively.}
\label{tab:results}
\vspace{0.2cm}
\def\arraystretch{1.2}
\begin{tabular}{lcc}
\hline\hline
\multicolumn{3}{c}{\em Spin-0}                                                                                      \\
\hline
\mX                  & ~~~400~\GeV        & ~~~2800~\GeV \\
NWA                  & 1.1~fb          & 0.03~fb \\
$\Gamma_X/\mX = 2\%$  & 2.5~fb          & 0.03~fb \\
$\Gamma_X/\mX = 6\%$  & 4.4~fb          & 0.03~fb \\
$\Gamma_X/\mX = 10\%$ & 8.3~fb          & 0.04~fb \\
\hline
\multicolumn{3}{c}{\em Spin-2}                                                                                      \\
\hline
\mG                & ~~~500~\GeV  & ~~~5000~\GeV \\
$\kMpl = 0.01$       & 1.9~fb  & 0.04~fb \\
$\kMpl = 0.05$       & 2.3~fb  & 0.04~fb \\
$\kMpl = 0.1$        & 3.2~fb  & 0.04~fb \\
\hline \hline
\end{tabular}
\end{table}

\begin{figure}
\centering
\subfloat[]{\includegraphics[width=.49\textwidth]{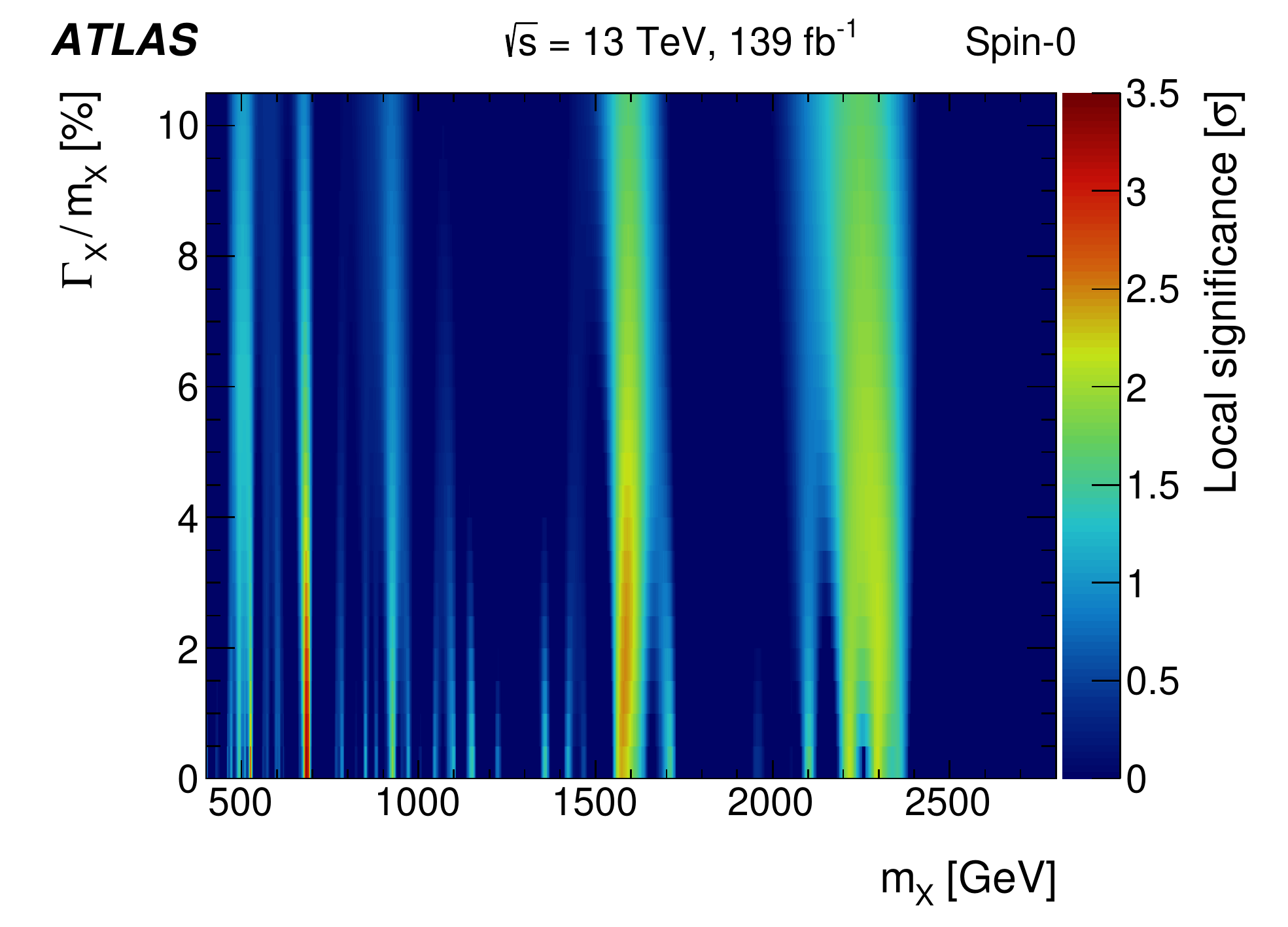}}
\subfloat[]{\includegraphics[width=.49\textwidth]{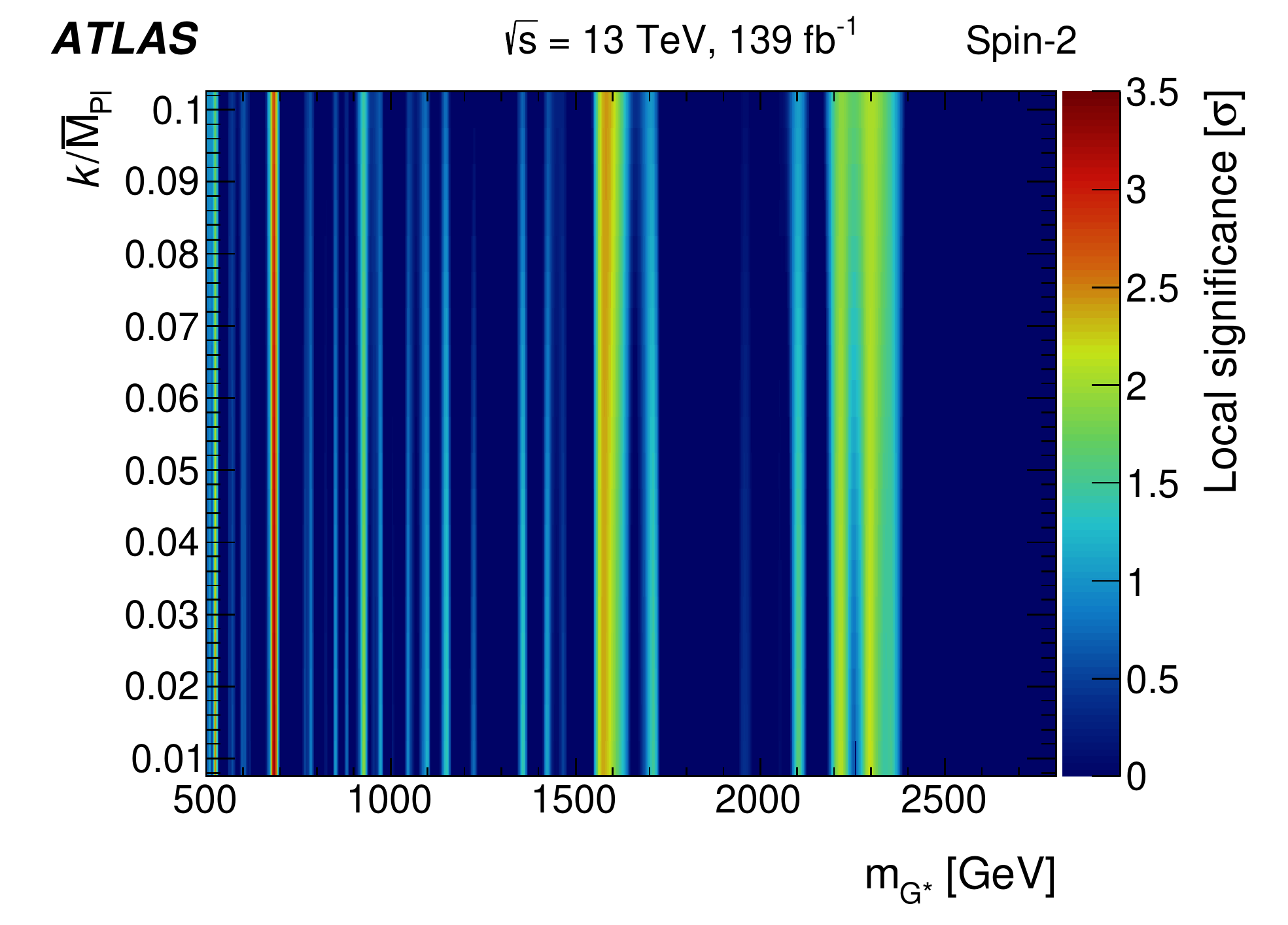}}
\caption{Probability of compatibility, in terms of local $p_{0}$ quantified in standard deviations $\sigma$, with the background-only hypothesis as a function of (a) the assumed signal mass \mX\ and relative width \GX/\mX\ for the spin-0 resonance search and (b) the assumed signal mass \mG\ and \kMpl\ for the spin-2 resonance search.}
\label{fig:2Dp0}
\end{figure}
 
\begin{figure}
\centering
\subfloat[]{\includegraphics[width=.49\textwidth]{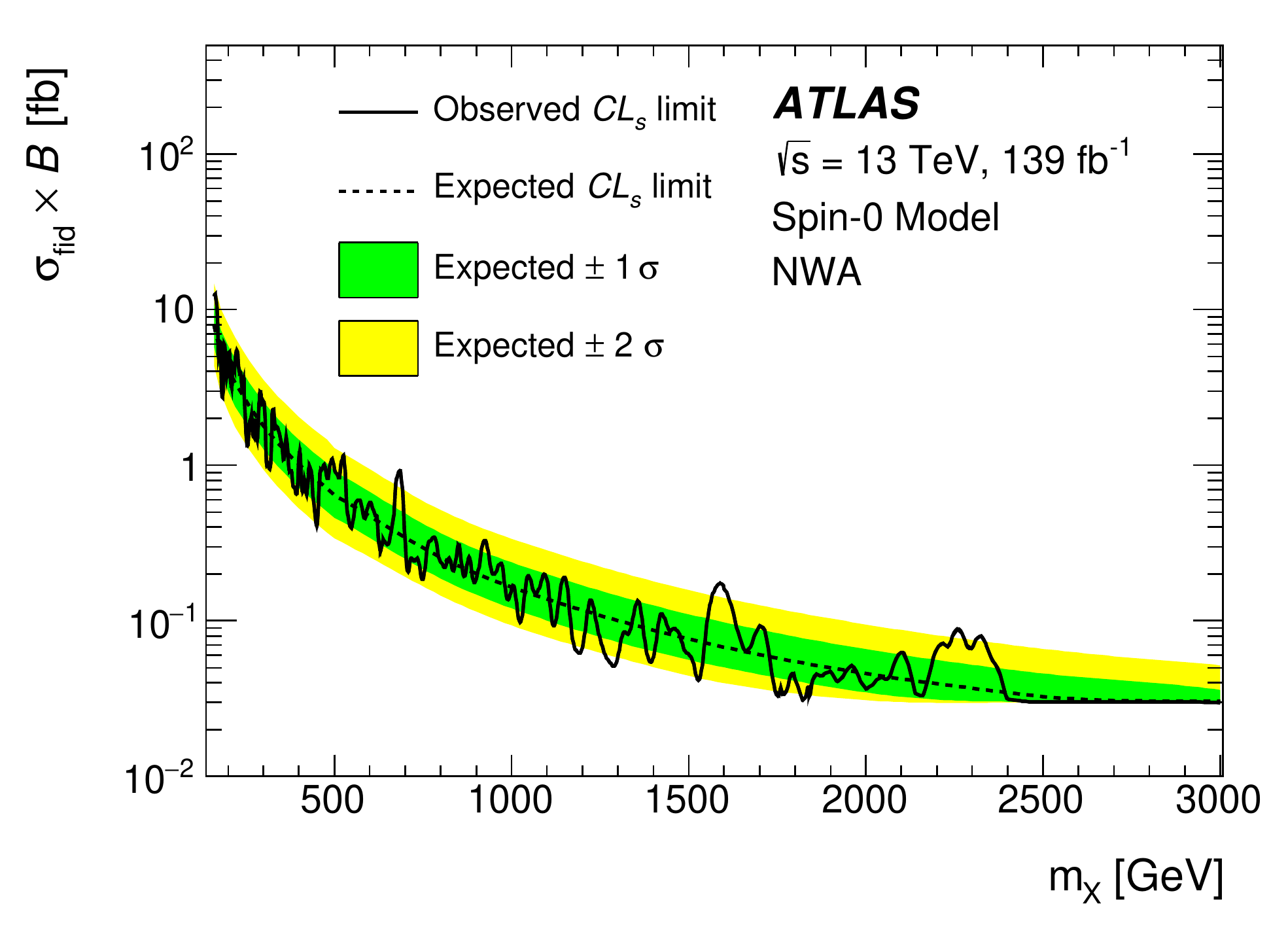}}
\subfloat[]{\includegraphics[width=.49\textwidth]{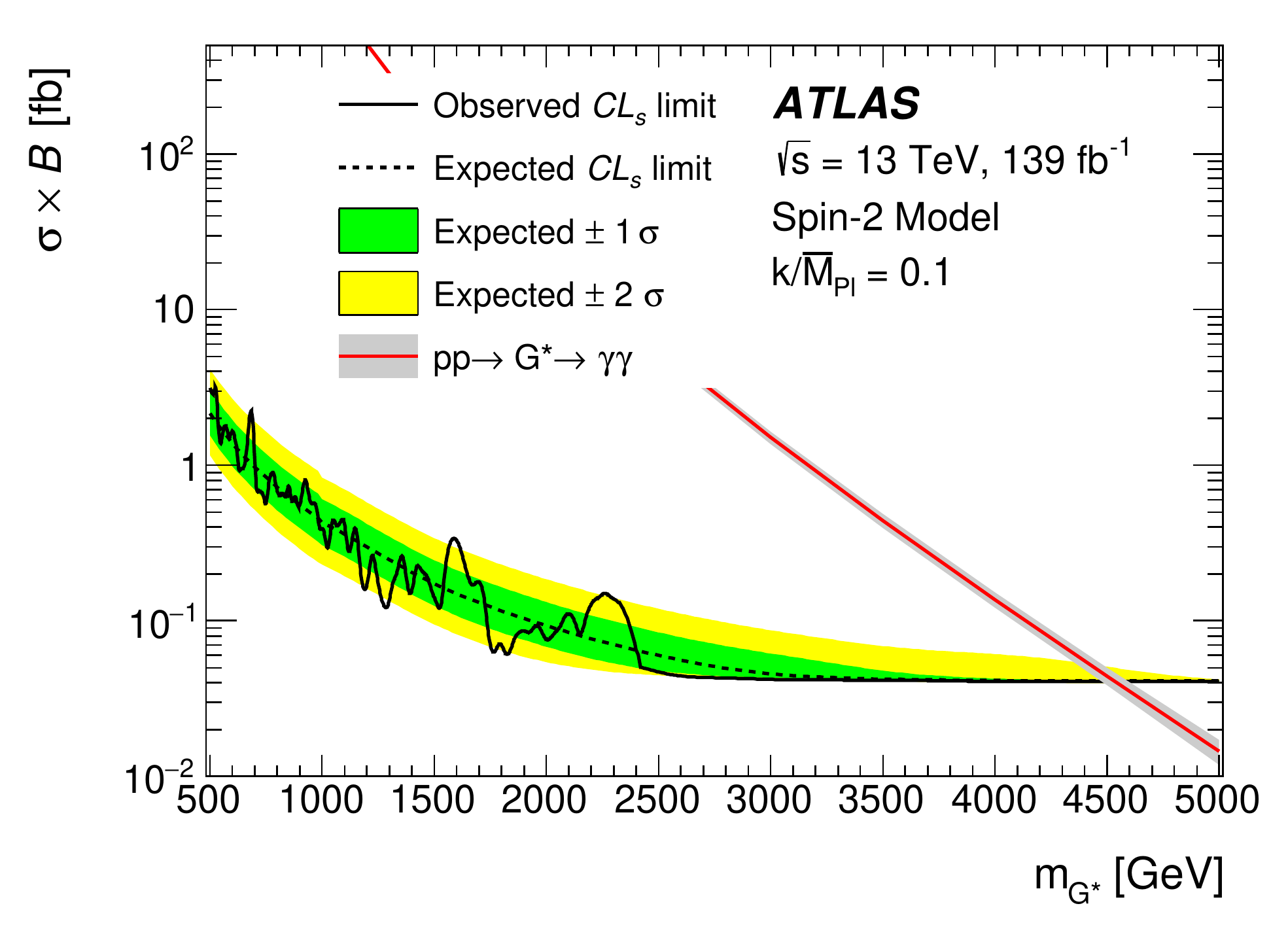}}
\caption{The expected and observed upper limits at 95\% CL on (a) the fiducial and (b) the total production cross-section times branching ratio to two photons of (a) a NWA spin-0 resonance as a function of its mass \mX\ and (b) the lightest KK graviton as a function of its mass \mG\ for $\kMpl = 0.1$. For masses greater than 1000~\GeV, pseudo-experiments are used to verify the expected and observed limits, and used in place of the asymptotic limit where differences are observed.
Cross-section predictions for the RS1 model are shown in (b), where the grey band illustrates the PDF uncertainty.
}
\label{fig:1Dlimits}
\end{figure}
 
\FloatBarrier

\ifnum\value{num}>0 {
\section{Conclusion}
} \else {}\fi

Searches for new resonances in high-mass diphoton final states with the ATLAS experiment at the LHC are presented. The searches use the full LHC Run 2 proton--proton collision dataset at a centre-of-mass energy of $\sqrt{s} = 13$~\TeV, corresponding to an integrated luminosity of 139~\ifb. The analyses are optimized to search for spin-0 resonances with masses above 200~\GeV\ and for spin-2 resonances predicted by the Randall--Sundrum model with masses above 500~\GeV.
 
The data are consistent with the Standard Model background expectation. In the spin-0 resonance search, the observed 95\% CL upper limits on the fiducial cross-section times branching ratio for a narrow-width signal range from 12.5~fb at 160~\GeV\ to about 0.03~fb at 2800~\GeV. In the spin-2 resonance search, the observed limits on the total cross-section times branching ratio for $\kMpl = 0.1$ range from 3.2~fb to about 0.04~fb for a graviton mass between 500 and 5000~\GeV, and the RS1 graviton is excluded for masses below 4500~\GeV\ for this coupling hypothesis.

 
\section*{Acknowledgements}
 

We thank CERN for the very successful operation of the LHC, as well as the
support staff from our institutions without whom ATLAS could not be
operated efficiently.
 
We acknowledge the support of
ANPCyT, Argentina;
YerPhI, Armenia;
ARC, Australia;
BMWFW and FWF, Austria;
ANAS, Azerbaijan;
SSTC, Belarus;
CNPq and FAPESP, Brazil;
NSERC, NRC and CFI, Canada;
CERN;
ANID, Chile;
CAS, MOST and NSFC, China;
Minciencias, Colombia;
MSMT CR, MPO CR and VSC CR, Czech Republic;
DNRF and DNSRC, Denmark;
IN2P3-CNRS and CEA-DRF/IRFU, France;
SRNSFG, Georgia;
BMBF, HGF and MPG, Germany;
GSRI, Greece;
RGC and Hong Kong SAR, China;
ISF and Benoziyo Center, Israel;
INFN, Italy;
MEXT and JSPS, Japan;
CNRST, Morocco;
NWO, Netherlands;
RCN, Norway;
MNiSW and NCN, Poland;
FCT, Portugal;
MNE/IFA, Romania;
JINR;
MES of Russia and NRC KI, Russian Federation;
MESTD, Serbia;
MSSR, Slovakia;
ARRS and MIZ\v{S}, Slovenia;
DSI/NRF, South Africa;
MICINN, Spain;
SRC and Wallenberg Foundation, Sweden;
SERI, SNSF and Cantons of Bern and Geneva, Switzerland;
MOST, Taiwan;
TAEK, Turkey;
STFC, United Kingdom;
DOE and NSF, United States of America.
In addition, individual groups and members have received support from
BCKDF, CANARIE, Compute Canada and CRC, Canada;
COST, ERC, ERDF, Horizon 2020 and Marie Sk{\l}odowska-Curie Actions, European Union;
Investissements d'Avenir Labex, Investissements d'Avenir Idex and ANR, France;
DFG and AvH Foundation, Germany;
Herakleitos, Thales and Aristeia programmes co-financed by EU-ESF and the Greek NSRF, Greece;
BSF-NSF and GIF, Israel;
Norwegian Financial Mechanism 2014-2021, Norway;
La Caixa Banking Foundation, CERCA Programme Generalitat de Catalunya and PROMETEO and GenT Programmes Generalitat Valenciana, Spain;
G\"{o}ran Gustafssons Stiftelse, Sweden;
The Royal Society and Leverhulme Trust, United Kingdom.
 
The crucial computing support from all WLCG partners is acknowledged gratefully, in particular from CERN, the ATLAS Tier-1 facilities at TRIUMF (Canada), NDGF (Denmark, Norway, Sweden), CC-IN2P3 (France), KIT/GridKA (Germany), INFN-CNAF (Italy), NL-T1 (Netherlands), PIC (Spain), ASGC (Taiwan), RAL (UK) and BNL (USA), the Tier-2 facilities worldwide and large non-WLCG resource providers. Major contributors of computing resources are listed in Ref.~\cite{ATL-SOFT-PUB-2021-003}.
 

\printbibliography

@article{Lee:1973iz,
    author = "Lee, T.D.",
    editor = "Feinberg, G.",
    title = "{A Theory of Spontaneous T Violation}",
    doi = "10.1103/PhysRevD.8.1226",
    journal = "Phys. Rev. D",
    volume = "8",
    pages = "1226--1239",
    year = "1973"
}

@article{Hill:1987ea,
    author = "Hill, A. and van der Bij, J.J.",
    title = "{Strongly interacting singlet - doublet Higgs model}",
    reportNumber = "UM-TH-86-19",
    doi = "10.1103/PhysRevD.36.3463",
    journal = "Phys. Rev. D",
    volume = "36",
    pages = "3463--3473",
    year = "1987"
}

@article{Veltman:1989vw,
    author = "Veltman, M.J.G. and Yndurain, F.J.",
    title = "{Radiative corrections to WW scattering}",
    reportNumber = "UM-TH-89-04",
    doi = "10.1016/0550-3213(89)90369-6",
    journal = "Nucl. Phys. B",
    volume = "325",
    pages = "1--17",
    year = "1989"
}

@article{Binoth:1996au,
    author = "Binoth, T. and van der Bij, J.J.",
    title = "{Influence of strongly coupled, hidden scalars on Higgs signals}",
    eprint = "hep-ph/9608245",
    archivePrefix = "arXiv",
    reportNumber = "FREIBURG-THEP-96-15",
    doi = "10.1007/s002880050442",
    journal = "Z. Phys. C",
    volume = "75",
    pages = "17--25",
    year = "1997"
}

@article{Schabinger:2005ei,
    author = "Schabinger, Robert M. and Wells, James D.",
    title = "{Minimal spontaneously broken hidden sector and its impact on Higgs boson physics at the CERN Large Hadron Collider}",
    eprint = "hep-ph/0509209",
    archivePrefix = "arXiv",
    doi = "10.1103/PhysRevD.72.093007",
    journal = "Phys. Rev. D",
    volume = "72",
    pages = "093007",
    year = "2005"
}

@article{Patt:2006fw,
    author = "Patt, Brian and Wilczek, Frank",
    title = "{Higgs-field Portal into Hidden Sectors}",
    eprint = "hep-ph/0605188",
    archivePrefix = "arXiv",
    reportNumber = "MIT-CTP-3745",
    month = "5",
    year = "2006"
}

@article{Pruna:2013bma,
    author = "Pruna, Giovanni Marco and Robens, Tania",
    title = "{Higgs singlet extension parameter space in the light of the LHC discovery}",
    eprint = "1303.1150",
    archivePrefix = "arXiv",
    primaryClass = "hep-ph",
    reportNumber = "PSI-PR-13-03",
    doi = "10.1103/PhysRevD.88.115012",
    journal = "Phys. Rev. D",
    volume = "88",
    number = "11",
    pages = "115012",
    year = "2013"
}

@book{Appelquist:1987nr,
    author = "Appelquist, T. and Chodos, A. and Freund, P.G.O.",
    title = "{Modern Kaluza-Klein theories}",
    publisher = "{Frontiers in Physics, Addison-Wesley}",
    year = "1987",
    volume = "65"
}

@article{Randall:1999ee,
    author = "Randall, Lisa and Sundrum, Raman",
    title = "{Large Mass Hierarchy from a Small Extra Dimension}",
    eprint = "hep-ph/9905221",
    archivePrefix = "arXiv",
    reportNumber = "MIT-CTP-2860, PUPT-1860, BUHEP-99-9",
    doi = "10.1103/PhysRevLett.83.3370",
    journal = "Phys. Rev. Lett.",
    volume = "83",
    pages = "3370--3373",
    year = "1999"
}

@article{Randall:1999vf,
    author = "Randall, Lisa and Sundrum, Raman",
    title = "{An Alternative to Compactification}",
    eprint = "hep-th/9906064",
    archivePrefix = "arXiv",
    reportNumber = "MIT-CTP-2874, PUPT-1867, BUHEP-99-13",
    doi = "10.1103/PhysRevLett.83.4690",
    journal = "Phys. Rev. Lett.",
    volume = "83",
    pages = "4690--4693",
    year = "1999"
}

@article{Davoudiasl:1999jd,
    author = "Davoudiasl, H. and Hewett, J.L. and Rizzo, T.G.",
    title = "{Phenomenology of the Randall-Sundrum Gauge Hierarchy Model}",
    eprint = "hep-ph/9909255",
    archivePrefix = "arXiv",
    reportNumber = "SLAC-PUB-8241",
    doi = "10.1103/PhysRevLett.84.2080",
    journal = "Phys. Rev. Lett.",
    volume = "84",
    pages = "2080",
    year = "2000"
}

@article{Catani:2001cc,
    author = "Catani, S. and Krauss, F. and Kuhn, R. and Webber, B.R.",
    title = "{QCD Matrix Elements + Parton Showers}",
    eprint = "hep-ph/0109231",
    archivePrefix = "arXiv",
    reportNumber = "CERN-TH-2000-367, CAVENDISH-HEP-00-03",
    doi = "10.1088/1126-6708/2001/11/063",
    journal = "JHEP",
    volume = "11",
    pages = "063",
    year = "2001"
}

@article{Hoeche:2011fd,
    author = "Hoeche, Stefan and Krauss, Frank and Schonherr, Marek and Siegert, Frank",
    title = "{A critical appraisal of NLO+PS matching methods}",
    eprint = "1111.1220",
    archivePrefix = "arXiv",
    primaryClass = "hep-ph",
    reportNumber = "SLAC-PUB-14661, IPPP-11-67, DCPT-11-134, LPN11-58, FR-PHENO-2011-019, MCNET-11-24",
    doi = "10.1007/JHEP09(2012)049",
    journal = "JHEP",
    volume = "09",
    pages = "049",
    year = "2012"
}

@Article{Bothmann:2019yzt,
    author = "Bothmann, Enrico and others",
    collaboration = "Sherpa",
    title = "{Event generation with Sherpa 2.2}",
    eprint = "1905.09127",
    archivePrefix = "arXiv",
    primaryClass = "hep-ph",
    reportNumber = "FERMILAB-PUB-19-218-T, SLAC-PUB-17433, IPPP/19/42, MCNET-19-11",
    doi = "10.21468/SciPostPhys.7.3.034",
    journal = "SciPost Phys.",
    volume = "7",
    number = "3",
    pages = "034",
    year = "2019"
}

@techreport{ATLAS-CONF-2019-021,
      title         = "{Luminosity determination in $pp$ collisions at
                       $\sqrt{s}=13$ TeV using the ATLAS detector at the LHC}",
      institution   = "CERN",
      collaboration = "ATLAS Collaboration",
      address       = "Geneva",
      number        = "ATLAS-CONF-2019-021",
      month         = "Jun",
      year          = "2019",
      reportNumber  = "ATLAS-CONF-2019-021",
      url           = "https://cds.cern.ch/record/2677054",
}

@Article{Cacciari:area,
author    = "Cacciari,
 Matteo and Salam, Gavin P. and Soyez, Gregory",
 journal   = "JHEP",
 volume    = "04",
 year      = "2008",
 pages     = "005",
 title     = "The catchment area of jets",
 doi       = "10.1088/1126-6708/2008/04/005",
 eprint    = "0802.1188",
 archivePrefix = "arXiv",
 primaryClass  =  "hep-ph",
 }

@Article{Cacciari:UE,
author    = "Cacciari, Matteo and Salam, Gavin P. and Sapeta, Sebastian",
title     = "{On the characterisation of the underlying event}",
journal   = "JHEP",
volume    = "04",
year      = "2010",
pages     = "065",
doi       = "10.1007/JHEP04(2010)065",
eprint    = "0912.4926",
archivePrefix = "arXiv",
primaryClass  =  "hep-ph",
}

@article{Lai:2010vv,
    author = "Lai, Hung-Liang and Guzzi, Marco and Huston, Joey and Li, Zhao and Nadolsky, Pavel M. and Pumplin, Jon and Yuan, C.-P.",
    title = "{New parton distributions for collider physics}",
    eprint = "1007.2241",
    archivePrefix = "arXiv",
    primaryClass = "hep-ph",
    reportNumber = "MSUHEP-100707, SMU-HEP-10-10",
    doi = "10.1103/PhysRevD.82.074024",
    journal = "Phys. Rev. D",
    volume = "82",
    pages = "074024",
    year = "2010"
}

@article{Bijnens:2001gh,
    author = "Bijnens, J. and Eerola, P. and Maul, M. and Mansson, A. and Sjostrand, T.",
    title = "{QCD signatures of narrow graviton resonances in hadron colliders}",
    eprint = "hep-ph/0101316",
    archivePrefix = "arXiv",
    reportNumber = "LU-TP-01-05",
    doi = "10.1016/S0370-2693(01)00238-6",
    journal = "Phys. Lett. B",
    volume = "503",
    pages = "341--348",
    year = "2001"
}

@article{Oreglia:1980cs,
    author = "Oreglia, M.",
    title = "{A Study of the Reactions $\psi^\prime \to \gamma \gamma \psi$}",
    journal = "SLAC-R-0236",
    month = "12",
    year = "1980",
    type = "Other thesis",
    url = "www.slac.stanford.edu/cgi-wrap/getdoc/slac-r-236.pdf"
}

@article{Edgar:2018irz,
    author = "Edgar, Ryan and Amidei, Dante and Grud, Christopher and Sekhon, Karishma",
    title = "{Functional Decomposition: A new method for search and limit setting}",
    eprint = "1805.04536",
    archivePrefix = "arXiv",
    primaryClass = "physics.data-an",
    month = "5",
    year = "2018"
}

@article{Junk:1999kv,
    author = "Junk, Thomas",
    title = "{Confidence level computation for combining searches with small statistics}",
    eprint = "hep-ex/9902006",
    archivePrefix = "arXiv",
    reportNumber = "CARLETON-OPAL-PHYS-99-01, CERN-EP-99-041",
    doi = "10.1016/S0168-9002(99)00498-2",
    journal = "Nucl. Instrum. Meth. A",
    volume = "434",
    pages = "435--443",
    year = "1999"
}

@article{Ball:2012cx,
      author         = "Ball, Richard D. and others",
      title          = "{Parton distributions with LHC data}",
      journal        = "Nucl. Phys. B",
      volume         = "867",
      year           = "2013",
      pages          = "244",
      doi            = "10.1016/j.nuclphysb.2012.10.003",
      eprint         = "1207.1303",
      archivePrefix  = "arXiv",
      primaryClass   = "hep-ph",
      reportNumber   = "EDINBURGH-2012-08, IFUM-FT-997, FR-PHENO-2012-014,
                        RWTH-TTK-12-25, CERN-PH-TH-2012-037, SFB-CPP-12-47\,
                        --CERN-PH-TH-2012-037",
      SLACcitation   = "%%CITATION = ARXIV:1207.1303;%%"
}

@article{Ball:2014uwa,
      author         = "Ball, Richard D. and others",
      title          = "{Parton distributions for the LHC run II}",
      collaboration  = "NNPDF",
      journal        = "JHEP",
      volume         = "04",
      year           = "2015",
      pages          = "040",
      doi            = "10.1007/JHEP04(2015)040",
      eprint         = "1410.8849",
      archivePrefix  = "arXiv",
      primaryClass   = "hep-ph",
      reportNumber   = "EDINBURGH-2014-15, IFUM-1034-FT, CERN-PH-TH-2013-253,
                        OUTP-14-11P, CAVENDISH-HEP-14-11",
      SLACcitation   = "%%CITATION = ARXIV:1410.8849;%%"
}

@article{Sjostrand:2014zea,
      author         = "Sj{\"o}strand, Torbj{\"o}rn and Ask, Stefan and Christiansen,
                        Jesper R. and Corke, Richard and Desai, Nishita and Ilten,
                        Philip and Mrenna, Stephen and Prestel, Stefan and
                        Rasmussen, Christine O. and Skands, Peter Z.",
      title          = "{An introduction to PYTHIA 8.2}",
      journal        = "Comput. Phys. Commun.",
      volume         = "191",
      year           = "2015",
      pages          = "159",
      doi            = "10.1016/j.cpc.2015.01.024",
      eprint         = "1410.3012",
      archivePrefix  = "arXiv",
      primaryClass   = "hep-ph",
      reportNumber   = "LU-TP-14-36, MCNET-14-22, CERN-PH-TH-2014-190,
                        FERMILAB-PUB-14-316-CD, DESY-14-178, SLAC-PUB-16122,
                        --FERMILAB-PUB-14-316-CD",
      SLACcitation   = "%%CITATION = ARXIV:1410.3012;%%"
}

@Article{Alwall:2014hca,
      author         = "Alwall, J. and Frederix, R. and Frixione, S. and Hirschi,
                        V. and Maltoni, F. and Mattelaer, O. and Shao, H. -S. and
                        Stelzer, T. and Torrielli, P. and Zaro, M.",
      title          = "{The automated computation of tree-level and
                        next-to-leading order differential cross sections, and
                        their matching to parton shower simulations}",
      journal        = "JHEP",
      volume         = "07",
      year           = "2014",
      pages          = "079",
      doi            = "10.1007/JHEP07(2014)079",
      eprint         = "1405.0301",
      archivePrefix  = "arXiv",
      primaryClass   = "hep-ph",
      reportNumber   = "CERN-PH-TH-2014-064, CP3-14-18, LPN14-066, MCNET-14-09,
                        ZU-TH-14-14",
      SLACcitation   = "%%CITATION = ARXIV:1405.0301;%%"
}

@Article{Nason:2004rx,
      author         = "Nason, Paolo",
      title          = "{A new method for combining NLO QCD with shower Monte Carlo algorithms}",
      journal        = "JHEP",
      volume         = "11",
      pages          = "040",
      doi            = "10.1088/1126-6708/2004/11/040",
      year           = "2004",
      eprint         = "hep-ph/0409146",
      archivePrefix  = "arXiv",
}

@Article{Frixione:2007vw,
      author         = "Frixione, Stefano and Nason, Paolo and Oleari, Carlo",
      title          = "{Matching NLO QCD computations with parton shower
                        simulations: the POWHEG method}",
      journal        = "JHEP",
      volume         = "11",
      pages          = "070",
      doi            = "10.1088/1126-6708/2007/11/070",
      year           = "2007",
      eprint         = "0709.2092",
      archivePrefix  = "arXiv",
      primaryClass   = "hep-ph",
}

@Article{Alioli:2010xd,
      author         = "Alioli, Simone and Nason, Paolo and Oleari, Carlo and Re,
                        Emanuele",
      title          = "{A general framework for implementing NLO calculations in
                        shower Monte Carlo programs: the POWHEG BOX}",
      journal        = "JHEP",
      volume         = "06",
      pages          = "043",
      doi            = "10.1007/JHEP06(2010)043",
      year           = "2010",
      eprint         = "1002.2581",
      archivePrefix  = "arXiv",
      primaryClass   = "hep-ph",
}

@Article{Gleisberg:2008ta,
      author         = "Gleisberg, T. and H{\"o}che, S. and Krauss, F. and Sch{\"o}nherr, M. and Schumann, S. and Siegert, F. and Winter J.",
      title          = "{Event generation with SHERPA 1.1}",
      journal        = "JHEP",
      volume         = "02",
      pages          = "007",
      doi            = "10.1088/1126-6708/2009/02/007",
      year           = "2009",
      eprint         = "0811.4622",
      archivePrefix  = "arXiv",
      primaryClass   = "hep-ph",
}

@Article{Hoeche:2009rj,
      author         = "H{\"o}che, Stefan and Krauss, Frank and Schumann, Steffen and Siegert, Frank",
      title          = "{QCD matrix elements and truncated showers}",
      journal        = "JHEP",
      volume         = "05",
      pages          = "053",
      doi            = "10.1088/1126-6708/2009/05/053",
      year           = "2009",
      eprint         = "0903.1219",
      archivePrefix  = "arXiv",
      primaryClass   = "hep-ph",
}

@article{Hoeche:2012yf,
      author         = "H{\"o}che, Stefan and Krauss, Frank and Sch{\"o}nherr, Marek and
                        Siegert, Frank",
      title          = "{QCD matrix elements + parton showers. The NLO case}",
      journal        = "JHEP",
      volume         = "04",
      year           = "2013",
      pages          = "027",
      doi            = "10.1007/JHEP04(2013)027",
      eprint         = "1207.5030",
      archivePrefix  = "arXiv",
      primaryClass   = "hep-ph",
      reportNumber   = "SLAC-PUB-15191, IPPP-12-52, DCPT-12-104, LPN12-081,
                        FR-PHENO-2012-017, MCNET-12-09, --FR-PHENO-2012-017",
      SLACcitation   = "%%CITATION = ARXIV:1207.5030;%%"
}

@article{Agostinelli:2002hh,
      author         = "{GEANT4 Collaboration} and Agostinelli, S. and others",
      title          = "{\textsc{Geant4} -- a simulation toolkit}",
      journal        = "Nucl. Instrum. Meth. A",
      volume         = "506",
      year           = "2003",
      pages          = "250",
      doi            = "10.1016/S0168-9002(03)01368-8",
      reportNumber   = "SLAC-PUB-9350, FERMILAB-PUB-03-339",
      SLACcitation   = "%%CITATION = NUIMA,A506,250;%%"
}

@Article{Read:2002hq,
    Author = {Read, Alexander L.},
    Journal = {J. Phys. G},
    Pages = {2693},
    doi = "10.1088/0954-3899/28/10/313",
    Title = "{Presentation of search results: the \(CL_S\) technique}",
    Volume = {28},
    Year = {2002}
}

@Article{Cowan:2010js,
   Author = {Cowan, Glen and Cranmer, Kyle and Gross, Eilam and Vitells, Ofer},
   Title = {{Asymptotic formulae for likelihood-based tests of new physics}},
   Journal = {Eur. Phys. J. C},
   Volume = {71},
   Year = {2011},
   Pages = {1554},
   doi = "10.1140/epjc/s10052-011-1554-0",
   Eprint = "1007.1727",
   Archiveprefix = {arXiv},
   Primaryclass = {physics.data-an},
   related = "Cowan:2010js-err",
   relatedstring  = "Erratum:",
}

@Article{PERF-2007-01,
    author         = "{ATLAS Collaboration}",
    title          = "{The ATLAS Experiment at the CERN Large Hadron Collider}",
    journal        = "JINST",
    volume         = "3",
    year           = "2008",
    pages          = "S08003",
    doi            = "10.1088/1748-0221/3/08/S08003",
    primaryClass   = "hep-ex",
}

@Article{SOFT-2010-01,
    author         = "{ATLAS Collaboration}",
    title          = "{The ATLAS Simulation Infrastructure}",
    journal        = "Eur. Phys. J. C",
    volume         = "70",
    year           = "2010",
    pages          = "823",
    doi            = "10.1140/epjc/s10052-010-1429-9",
    eprint         = "1005.4568",
    archivePrefix  = "arXiv",
    primaryClass   = "physics.ins-det",
}

@Article{STDM-2012-05,
    author         = "{ATLAS Collaboration}",
    title          = "{Measurement of isolated-photon pair production in \(pp\) collisions at \(\sqrt{s} = 7\,\text{TeV}\) with the ATLAS detector}",
    journal        = "JHEP",
    volume         = "01",
    year           = "2013",
    pages          = "086",
    doi            = "10.1007/JHEP01(2013)086",
    reportNumber   = "CERN-PH-EP-2012-300",
    eprint         = "1211.1913",
    archivePrefix  = "arXiv",
    primaryClass   = "hep-ex",
}

@Article{STDM-2012-23,
    author         = "{ATLAS Collaboration}",
    title          = "{Measurement of the \(Z/\gamma^*\) boson transverse momentum distribution in \(pp\) collisions at \(\sqrt{s} = 7\,\text{TeV}\) with the ATLAS detector}",
    journal        = "JHEP",
    volume         = "09",
    year           = "2014",
    pages          = "145",
    doi            = "10.1007/JHEP09(2014)145",
    reportNumber   = "CERN-PH-EP-2014-075",
    eprint         = "1406.3660",
    archivePrefix  = "arXiv",
    primaryClass   = "hep-ex",
}

@Article{HIGG-2013-08,
    author         = "{ATLAS Collaboration}",
    title          = "{Measurement of Higgs boson production in the diphoton decay channel in \(pp\) collisions at center-of-mass energies of \(7\) and \(8\,\text{TeV}\) with the ATLAS detector}",
    journal        = "Phys. Rev. D",
    volume         = "90",
    year           = "2014",
    pages          = "112015",
    doi            = "10.1103/PhysRevD.90.112015",
    reportNumber   = "CERN-PH-EP-2014-198",
    eprint         = "1408.7084",
    archivePrefix  = "arXiv",
    primaryClass   = "hep-ex",
}

@Article{HIGG-2014-04,
    author         = "{ATLAS Collaboration}",
    title          = "{Search for Scalar Diphoton Resonances in the Mass Range 65--600\,GeV with the ATLAS Detector in \(pp\) Collision Data at \(\sqrt{s} = 8\,\text{TeV}\)}",
    journal        = "Phys. Rev. Lett.",
    volume         = "113",
    year           = "2014",
    pages          = "171801",
    doi            = "10.1103/PhysRevLett.113.171801",
    reportNumber   = "CERN-PH-EP-2014-142",
    eprint         = "1407.6583",
    archivePrefix  = "arXiv",
    primaryClass   = "hep-ex",
}

@Article{PERF-2014-07,
    author         = "{ATLAS Collaboration}",
    title          = "{Topological cell clustering in the ATLAS calorimeters and its performance in LHC Run~1}",
    journal        = "Eur. Phys. J. C",
    volume         = "77",
    year           = "2017",
    pages          = "490",
    doi            = "10.1140/epjc/s10052-017-5004-5",
    reportNumber   = "CERN-PH-EP-2015-304",
    eprint         = "1603.02934",
    archivePrefix  = "arXiv",
    primaryClass   = "hep-ex",
}

@Article{HIGG-2016-08,
    author         = "{ATLAS Collaboration}",
    title          = "{Search for resonances in diphoton events at \(\sqrt{s} = 13\,\text{TeV}\) with the ATLAS detector}",
    journal        = "JHEP",
    volume         = "09",
    year           = "2016",
    pages          = "001",
    doi            = "10.1007/JHEP09(2016)001",
    reportNumber   = "CERN-EP-2016-120",
    eprint         = "1606.03833",
    archivePrefix  = "arXiv",
    primaryClass   = "hep-ex",
}

@Article{HIGG-2016-17,
    author         = "{ATLAS Collaboration}",
    title          = "{Search for new phenomena in high-mass diphoton final states using \(37\,\text{fb}^{-1}\) of proton--proton collisions collected at \(\sqrt{s} = 13\,\text{TeV}\) with the ATLAS detector}",
    journal        = "Phys. Lett. B",
    volume         = "775",
    year           = "2017",
    pages          = "105",
    doi            = "10.1016/j.physletb.2017.10.039",
    reportNumber   = "CERN-EP-2017-132",
    eprint         = "1707.04147",
    archivePrefix  = "arXiv",
    primaryClass   = "hep-ex",
}

@Article{EGAM-2018-01,
    author         = "{ATLAS Collaboration}",
    title          = "{Electron and photon performance measurements with the ATLAS detector using the 2015--2017 LHC proton--proton collision data}",
    journal        = "JINST",
    volume         = "14",
    year           = "2019",
    pages          = "P12006",
    doi            = "10.1088/1748-0221/14/12/P12006",
    reportNumber   = "CERN-EP-2019-145",
    eprint         = "1908.00005",
    archivePrefix  = "arXiv",
    primaryClass   = "hep-ex",
}

@Article{HIGG-2018-27,
    author         = "{ATLAS Collaboration}",
    title          = "{Search for resonances decaying into photon pairs in \(139\,\text{fb}^{-1}\) of \(pp\) collisions at \(\sqrt{s} = 13\,\text{TeV}\) with the ATLAS detector}",
    year           = "2021",
    reportNumber   = "CERN-EP-2020-248",
    eprint         = "2102.13405",
    archivePrefix  = "arXiv",
    primaryClass   = "hep-ex",
}

@Article{PIX-2018-001,
    author         = "Abbott, B. and others",
    title          = "{Production and integration of the ATLAS Insertable B-Layer}",
    journal        = "JINST",
    volume         = "13",
    year           = "2018",
    pages          = "T05008",
    doi            = "10.1088/1748-0221/13/05/T05008",
    eprint         = "1803.00844",
    archivePrefix  = "arXiv",
    primaryClass   = "physics.ins-det",
}

@Article{TRIG-2018-05,
    author         = "{ATLAS Collaboration}",
    title          = "{Performance of electron and photon triggers in ATLAS during LHC Run~2}",
    journal        = "Eur. Phys. J. C",
    volume         = "80",
    year           = "2020",
    pages          = "47",
    doi            = "10.1140/epjc/s10052-019-7500-2",
    reportNumber   = "CERN-EP-2019-169",
    eprint         = "1909.00761",
    archivePrefix  = "arXiv",
    primaryClass   = "hep-ex",
}

@Booklet{ATL-SOFT-PUB-2021-003,
    author         = "{ATLAS Collaboration}",
    title          = "{ATLAS Computing Acknowledgements}",
    howpublished   = "{ATL-SOFT-PUB-2021-003}",
    url            = "https://cds.cern.ch/record/2776662",
}

@Report{ATLAS-TDR-2010-19,
    author         = "{ATLAS Collaboration}",
    title          = "{ATLAS Insertable B-Layer Technical Design Report}",
    type           = "ATLAS-TDR-19; CERN-LHCC-2010-013",
    year           = "2010",
    url            = "https://cds.cern.ch/record/1291633",
    related        = "ATLAS-TDR-2010-19-addm",
    relatedstring  = "Addendum:",
}

@Article{CMS-EXO-17-017,
    author         = "{CMS Collaboration}",
    title          = "{Search for physics beyond the standard model in high-mass diphoton events from proton--proton collisions at \(\sqrt{s} = 13\,\text{TeV}\)}",
    journal        = "Phys. Rev. D",
    volume         = "98",
    year           = "2018",
    pages          = "092001",
    doi            = "10.1103/PhysRevD.98.092001",
    reportNumber   = "CERN-EP-2018-219",
    eprint         = "1809.00327",
    archivePrefix  = "arXiv",
    primaryClass   = "hep-ex",
}

@Booklet{ATL-PHYS-PUB-2010-013,
    author         = "{ATLAS Collaboration}",
    title          = "{The simulation principle and performance of the ATLAS fast calorimeter simulation FastCaloSim}",
    howpublished   = "{ATL-PHYS-PUB-2010-013}",
    url            = "https://cds.cern.ch/record/1300517",
    year           = "2010",
}

@Booklet{ATL-PHYS-PUB-2014-021,
    author         = "{ATLAS Collaboration}",
    title          = "{ATLAS Pythia~8 tunes to \(7~\text{TeV}\) data}",
    howpublished   = "{ATL-PHYS-PUB-2014-021}",
    url            = "https://cds.cern.ch/record/1966419",
    year           = "2014",
}

\clearpage 
 
\begin{flushleft}
\hypersetup{urlcolor=black}
{\Large The ATLAS Collaboration}

\bigskip

\AtlasOrcid[0000-0002-6665-4934]{G.~Aad}$^\textrm{\scriptsize 102}$,    
\AtlasOrcid[0000-0002-5888-2734]{B.~Abbott}$^\textrm{\scriptsize 129}$,    
\AtlasOrcid[0000-0002-7248-3203]{D.C.~Abbott}$^\textrm{\scriptsize 103}$,    
\AtlasOrcid[0000-0002-2788-3822]{A.~Abed~Abud}$^\textrm{\scriptsize 36}$,    
\AtlasOrcid[0000-0002-1002-1652]{K.~Abeling}$^\textrm{\scriptsize 53}$,    
\AtlasOrcid[0000-0002-2987-4006]{D.K.~Abhayasinghe}$^\textrm{\scriptsize 94}$,    
\AtlasOrcid[0000-0002-8496-9294]{S.H.~Abidi}$^\textrm{\scriptsize 168}$,    
\AtlasOrcid[0000-0002-8279-9324]{O.S.~AbouZeid}$^\textrm{\scriptsize 40}$,    
\AtlasOrcid{N.L.~Abraham}$^\textrm{\scriptsize 157}$,    
\AtlasOrcid[0000-0001-5329-6640]{H.~Abramowicz}$^\textrm{\scriptsize 162}$
% ...
\bigskip
\\

% ...
$^{36}$CERN, Geneva; Switzerland.\\
% ...
$^{40}$Niels Bohr Institute, University of Copenhagen, Copenhagen; Denmark.\\
% ...
$^{53}$II. Physikalisches Institut, Georg-August-Universit\"{a}t G\"ottingen, G\"ottingen; Germany.\\
% ...
$^{94}$Department of Physics, Royal Holloway University of London, Egham; United Kingdom.\\
% ...
$^{102}$CPPM, Aix-Marseille Universit\'e, CNRS/IN2P3, Marseille; France.\\
$^{103}$Department of Physics, University of Massachusetts, Amherst MA; United States of America.\\
% ...
$^{129}$Homer L. Dodge Department of Physics and Astronomy, University of Oklahoma, Norman OK; United States of America.\\
% ...
$^{157}$Department of Physics and Astronomy, University of Sussex, Brighton; United Kingdom.\\
% ...
$^{162}$Raymond and Beverly Sackler School of Physics and Astronomy, Tel Aviv University, Tel Aviv; Israel.\\
% ...
$^{168}$Department of Physics, University of Toronto, Toronto ON; Canada.\\
% ...

\end{flushleft}


\end{document}